\documentclass[twocolumn,showpacs,preprintnumbers,amsmath,amssymb]{revtex4}
\usepackage{graphicx}
\begin{document}
\title{Momentum transfer using chirped standing wave fields:
Bragg scattering}
\author{Vladimir~S.~Malinovsky and Paul~R.~Berman}
\address{Michigan Center for Theoretical Physics\\
$\&$ FOCUS Center, Department of Physics,\\
University of Michigan, Ann Arbor, MI 48109-1120}
\begin{abstract}
\noindent 
We consider momentum transfer using frequency-chirped standing
wave fields. Novel atom-beam splitter and mirror schemes based on 
Bragg scattering are presented. It is shown that a predetermined 
number of photon momenta can be transferred to the atoms in a 
single interaction zone.
\end{abstract}
 
\maketitle
Atom optics has experienced rapid advances in recent years. 
Applications of atom optics to inertial sensing~\cite{GusPRL97}, 
atom holography~\cite{ZobPRA99,FujPRL00} and certain schemes for 
quantum computing~\cite{BrePRL99,JakPRL99} can benefit substantially 
from the ability to manipulate atomic motion in a controllable way. 
There are a number of theoretical and experimental studies devoted 
to this problem~\cite{BerAI,AdaPRS94,MetPRS94,WalPRS95}.

The underlying physical mechanism responsible for optical control 
of atomic motion is an exchange of momentum between the atoms and 
the fields. Momentum exchange can be used as the basis of practical 
devices, such as atom mirrors and atom beam splitters that are 
essential elements of an atom interferometer. Optical $\pi /2$ and 
$\pi $ pulses have been used to create and deflect coherent 
superposition states involving different ground state
sublevels~\cite{KasPRL91,McgPRL00}. These experiments require one 
to control pulse power or duration to a fairly high precision. 
Alternative methods for producing large angle beam splitters involve 
the use of magneto-optical potentials, bichromatic forces, and 
strong standing wave fields~\cite{DubPRA01}. Bragg scattering 
involving multiphoton transitions~\cite{BerPRA97} can also be used 
to produce large angle splitting, but the power requirements 
increase and the resolution decreases with increasing order of
the transitions. 

To avoid the difficulties involved with pulses having specified 
areas, Rapid Adiabatic Passage methods have been 
proposed~\cite{MarPRL91,BanPRA93}. Population transfer by Stimulated 
Raman Adiabatic Passage (STIRAP) using delayed laser pulses was 
first observed by Bergmann and co-workers~\cite{GauCPL88}. The use 
of this method to create an atom beam splitter was
proposed in~\cite{MarPRL91}. It is worthwhile to mention that the 
adiabatic passage method is robust against changes in pulse 
parameters. Furthermore, momentum transfer based on the STIRAP 
technique~\cite{MarPRL91,PilPRA93,WeiPRL94,LawPRL94,GolPRL94} cannot 
be degraded as a result of spontaneous decay, as the excited states 
are never populated.

In this paper we propose schemes to create a 50/50 beam splitter and 
a mirror using frequency-chirped standing waves and adiabatic rapid 
passage. Both schemes are based on high order Bragg scattering and 
utilize an off-resonant interaction of two-level atoms with 
frequency-chirped standing-wave fields. The off-resonant atom-field 
interaction allows to one to avoid populating excited states; atoms 
remain in their ground state during the entire evolution of the 
atom-field interaction and spontaneous emission plays a negligible 
role. We use chirped pulses to produce efficient momentum transfer 
sequentially to states having momentum $\pm n2\hbar k$,
where $k$ is the propagation vector of one of the fields and $n$ is 
a positive integer. The chirp rate and pulse duration are used to 
control the final target state. This work is complementary to that 
involving the use of adiabatic rapid passage to accelerate atoms 
that are trapped in optical potentials~\cite{raizen}.

We consider first an atom mirror, which allows us to understand the 
dynamics of momentum transfer. The problem is closely related to 
population transfer in multilevel systems using Raman chirped 
adiabatic passage~\cite{Che95R3417,ChaCPL373}, but the level scheme 
differs somewhat (see Fig.~3 of ref.~\cite{BerPRA97}), in that we 
have a doubly degenerate ladder of Bragg states. The origin of this 
spectrum is discussed below.

An atomic beam having longitudinal velocity ${\bf u}$ in the 
${\bf \hat{x}}$ direction crosses a field region in which two 
optical fields counterpropagate in the ${\bf \hat{z}}$ direction. 
The longitudinal motion is treated classically, but the transverse 
motion (parallel to the field propagation vectors) is quantized. 
The fields couple the atomic ground state $|1\rangle $ to an excited 
state $|2\rangle $ having energy $E_{21}$. The Hamiltonian 
describing the atom field interaction is 
\begin{equation}
\begin{array}{lll}
\widehat{H}=\frac{\widehat{p}^{2}}{2m}+E_{21}|2\rangle 
\langle 2|-\left[ \Bigl(\mu _{12}E_{k}(t)\cos (\omega _{1}(t)+kz)\right.  \\ 
\left. {}\right.  \\ 
+\left. \mu _{12}E_{-k}(t)\cos (\omega _{2}(t)-kz)\Bigr)|1\rangle \langle
2|+h.c.\right] , 
\end{array}
\label{eq:TDSER} 
\end{equation}
\noindent 
where $\widehat{p}$ is the center-of-mass momentum operator of the
atom in the ${\bf \hat{z}}$ direction, $\mu _{12}$ is a dipole moment 
matrix element and $m$ is the atomic mass. The laser fields have wave 
vectors $\pm k {\bf \hat{z}}$, pulse envelopes $E_{\pm k}(t)$ as seen 
in the atomic rest frame moving with velocity ${\bf u}$ relative to 
the laboratory frame, and time-dependent phases 
$\omega _{1}(t)=\omega _{0}t+\delta _{1}(t)$, 
$\omega_{2}(t)=\omega _{0}t+\delta _{2}(t)+\delta _{0}t$, 
where $\omega _{0}$ is a central frequency.

Assuming that the detuning, $\Delta = E_{21}/\hbar - \omega_{0}$, 
is large compared with $\left| d\left[\omega_{2}(t) - \omega_{1}(t)\right] 
/dt\right|$, Rabi frequencies $\left| \Omega_{\pm k}(t)\right| = 
|\mu_{12}E_{\pm k}(t)/\hbar|$, and the transverse kinetic energy term 
divided by $\hbar$, we make the rotating wave approximation and 
adiabatically eliminate the excited state amplitude. The equation 
of motion for the ground state wave function in the momentum 
representation takes the following form 
\begin{equation}
\begin{array}{lll}
i\dot{a}(p,t) = \frac{p^{2}}{2m \hbar} a(p,t) - \Omega_{e}(t) 
\left[ \exp[i\phi(t)] a(p - 2\hbar k,t)\right. \\ 
\left. {}\right. \\ 
\left. + \exp [-i\phi(t)] a(p + 2\hbar k,t)\right] ,
\end{array}
\label{eq:momentspace}
\end{equation}
\noindent 
where $\Omega_{e}(t)=\Omega_{k}(t)\Omega_{-k}(t)/(4\Delta )$ is an
effective Rabi frequency, $\phi(t) = \delta_{1}(t) - \delta_{2}(t) -
\delta_{0}t$. We have omitted a factor depending on the light shift 
$\Omega_{\Delta}(t) = (\Omega_{k}^{2}(t) + \Omega_{-k}^{2}(t))/(4\Delta )$ 
in the Eq.~(\ref{eq:momentspace}) since this corresponds simply to a
redefinition of the ground state energy.

From Eq.~(\ref{eq:momentspace}) it is clear that states with momentum 
$p$ couple only to the neighboring states $p\pm 2\hbar k$. Using the 
initial condition $a(p,t=0)=\delta(p)$ one can set 
$$
a(p,t)=\sum_{n=-\infty }^{\infty }a_{n}(t)\delta (p-2n\hbar k) 
\exp[i n \phi(t)]
$$
and write Eq.~(\ref{eq:momentspace}) in matrix form for the amplitudes 
$a_{n}(t)$ with the Hamiltonian 
\begin{equation}
H(t) = \left\{ 
\begin{array}{cccccc}
\ddots & \ddots & 0 &  &  &  \\ 
\ddots & E_{-1}(t)/\hbar & - \Omega_{-n}(t) & 0 &  &  \\ 
0 & - \Omega_{-n}^{\ast}(t) & E_{0}(t)/\hbar & - \Omega_{+n}(t) & 0 &  \\ 
& 0 & - \Omega_{+n}^{\ast}(t) & E_{+1}(t)/\hbar & \ddots &  \\ 
&  & 0 & \ddots & \ddots & 
\end{array}
\right\} \,,  \label{eq:Hmatrix}
\end{equation}
where $\Omega_{-n}(t)=\Omega_{+n}(t)=\Omega_{e}(t)$. 
The quasi-energies in Eq.~(\ref{eq:Hmatrix}) are given by 
$E_{n}(t) = \hbar \left\{ n^2 \omega_k +  n \dot{\phi}(t)\right\}$, 
where $\omega_{k}= 2 \hbar k^{2} / m$ is a (two-photon) recoil 
frequency, and $n=0, \pm 1, \pm 2, \cdots$. This expression for 
the quasienergies is easy to understand. The Bragg levels have 
energy $E_{n}= \hbar n^{2} \omega_k$ and the ground state ($n=0$) 
is coupled to state $n$ by an $n$-(two)-photon process having 
effective frequency $\dot{\phi}(t)$. Thus the field energy is 
lowered by $n \dot{\phi}(t)$ when the atom is excited to
state $n$ and this loss is reflected in the quasienergies of the 
atom+field. The Hamiltonian (\ref{eq:Hmatrix}) describe dynamics 
in momentum space and $|a_{n}(t)|^{2}$ are the probability for an 
atom to have momentum $2n \hbar k$.
\begin{figure}[h]
\includegraphics[height=8.5cm,width=6cm,angle=270]{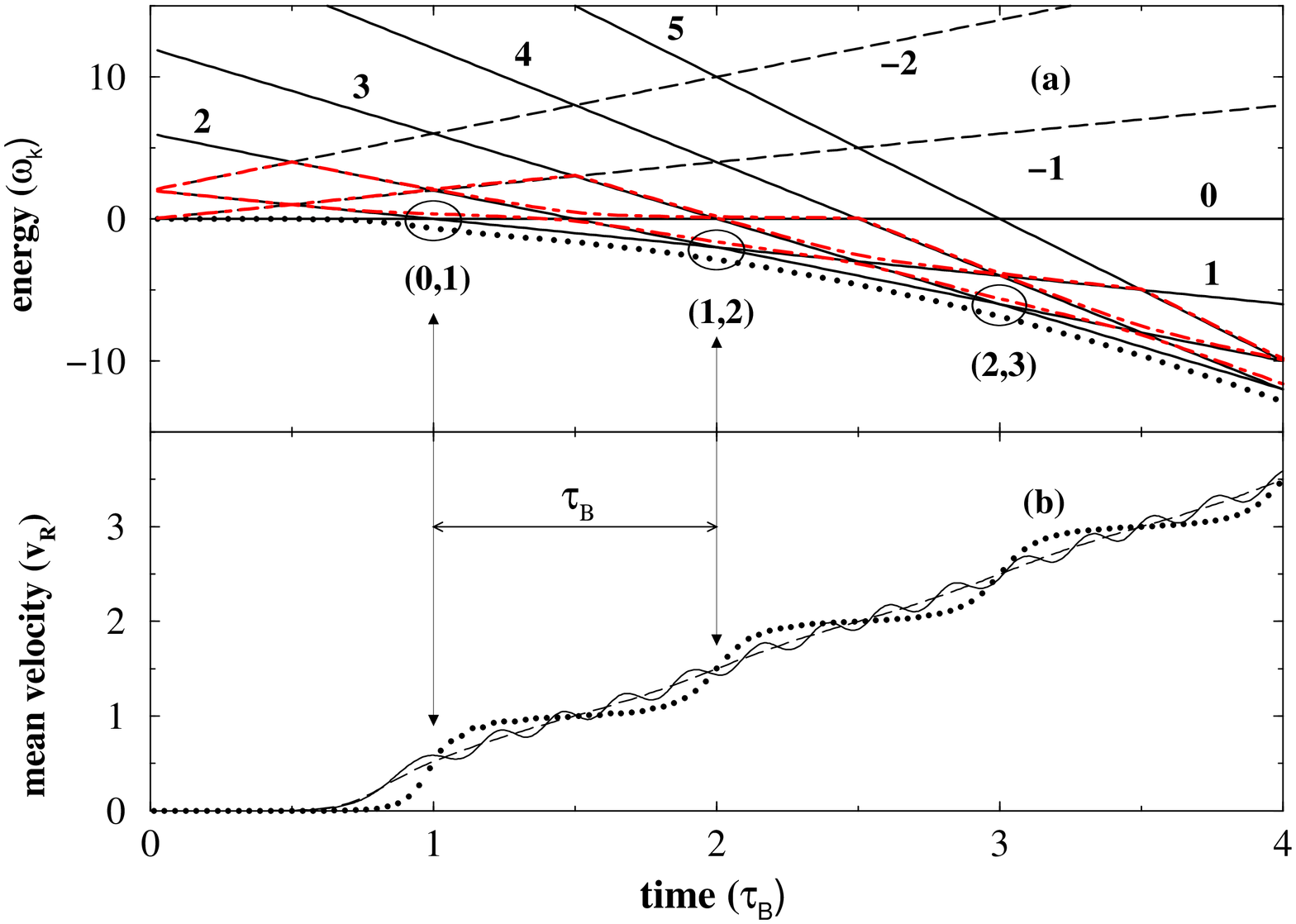}
\noindent 
\caption{(a) Dressed states as a function of time for the case of an 
atomic mirror. Dark solid lines show diabatic states of the positive 
branch, dashed lines show diabatic states of the negative branch. 
The dotted line shows the lowest adiabatic state. Several of the 
closest adiabatic states (dot-dashed lines) and a few avoided 
crossings are shown. 
(b) Mean velocity vs time. Dotted line shows Bloch oscillations, 
$\protect\alpha =0.01\omega_k^2$, $\Omega_{0}=0.15\omega_k$.
Dashed ($\protect\alpha =0.01\omega_k^2$) and solid 
($\protect\alpha =0.1\omega_k^2$) lines show two different 
regimes at $\Omega _{0}=0.7\omega_k$.}
\end{figure}

In the case of an linear-chirped standing wave field 
$\dot{\phi}(t)=\alpha (t-t_{c})-\delta _{0}$ , 
where $\alpha $ is the chirp rate and $t_{c}$ is a constant. 
It is clear that in a diabatic representation there are many 
sequential crossings between the $E_{n}(t)$. They become avoided 
crossings owing to the interaction with the laser fields 
(see Fig.~1). The resonances can be viewed as sequential two-photon 
Bragg resonances equally spaced in time with period 
$\Delta t=2 \omega_{k}/\alpha$.
The dotted line in Fig.~1(a) represents the lowest adiabatic
state. This state correlates with the zero momentum state as the 
pulse arrives and with the target state following the 
pulse. When adiabatic conditions are satisfied for all sequential 
crossings, atoms remain in this instantaneous eigenstate which 
evolves into the state having momentum $2n\hbar k$. 

Another way to consider momentum transfer in an optical lattice is to
describe motion of an atom in the periodic potential under the influence a
constant force using the Bloch formalism~\cite{PeiPRA2989}. In this
picture one can observe Bloch oscillations of the mean atomic
velocity as shown in Fig.~1(b) by dotted line. The adiabatic sequential
momentum transfer mentioned above corresponds to Bloch oscillations in the
lowest band. The amplitude of these oscillations is suppressed with
increasing Rabi frequency (dashed line). To satisfy adiabatic conditions 
one must use a very small chirp rate, $\alpha ,$ resulting in a long period 
for Bloch oscillations, $\tau _{B}=2\omega _{k}/\alpha $, and, consequently, 
in a long time for momentum transfer. We show below a possibility to reduce
considerably the total absolute time of momentum transfer by increasing
the chirp rate. An increasing chirp rate breaks adiabaticity at the time of 
the first avoided crossing, $(0,1)$ (see Fig.~1(a)) and nonadiabatic 
couplings became important in this regime. However, one can still have 
efficient transfer to the target state provided $\alpha $ is not too large. 
The oscillations in the solid curve of Fig. 1 (b) are evidence for 
nonadiabatic effects; nevertheless, the transfer to the target state in 
nearly 100 \% for these parameters, as is shown below. 

According to the structure of Eq.~(\ref{eq:Hmatrix}) the initial state, 
$a_{0}(t)$, is connected to the $\pm 1$ states. From the expression for 
the quasi-energies, $E_n(t)$, it follows that a positive chirp will 
sequentially bring the effective frequency into resonance with transition 
frequencies between states of the positive branch while states of the 
negative branch will be shifted more and more from resonance 
(see Fig.~1(a)). Consequently for successful momentum transfer we have 
to make sure that transient population of the $-2 \hbar k$ state due to 
the off-resonant interaction with the laser field is minimal. As we show 
in the example below this can be accomplished by adjusting a switching-on 
stage of the pulse and the chirp rate.
\begin{figure}[h]
\includegraphics[height=9cm,width=8cm]{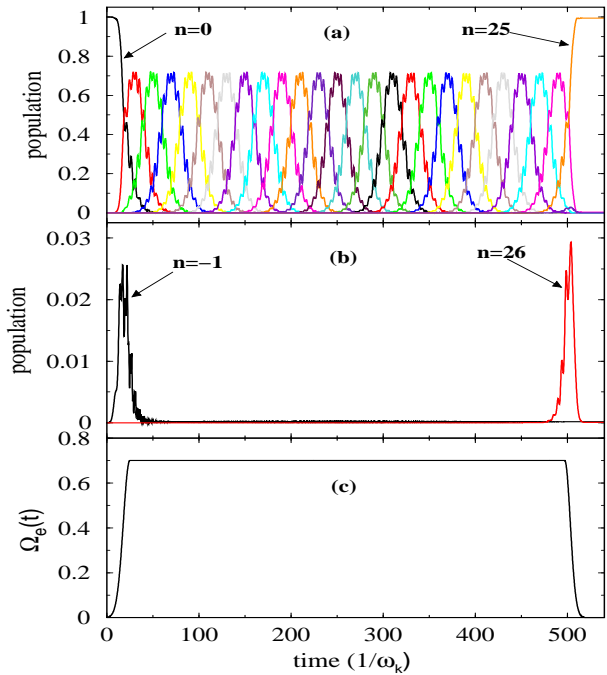}
\noindent 
\caption{Population dynamics for an atomic mirror. We choose $t_{c}=10/
\protect\omega _{k}$, $\protect\alpha =0.1\protect\omega _{k}^{2}$. (a)
Solid lines show the population flow to the target state $n=25$. (b)
Population of the $n=-1$ and of the $n=26$ states. (c) Shape of the 
laser pulse.}
\end{figure}

Figure~2 shows the momentum transfer to the $50\hbar k$ state for the
initial condition $a_{0}(t=0)~=~1$. This figure demonstrates almost 
100\% efficiency of adiabatic momentum transfer and corresponds to an 
atom mirror. As a target we chose the $50\hbar k$ state; however, 
in principle, there is no limit to the number of the momentum quanta 
$2\hbar k$ which can be transferred to the atoms. As long as we 
approximately satisfy an adiabatic condition at the beginning 
of the pulse, when the field is not so strong, momentum transfer 
is nearly 100\% efficient for the sequential Bragg resonances. 
The pulse duration is the parameter that determines how many
transitions take place. After a target state is chosen the turn-off 
stage of the pulse can be adjusted as at the beginning of the pulse 
to avoid population of higher states. 

There is enough freedom to change the time interval between sequential
momentum transfers by adjusting the chirp rate. In the case of the atom
mirror (see Fig.~2) the time interval between crossings, $\Delta t =2
\omega_{k}/\alpha = 20$, in units of $\omega_{k}^{-1}$. At the same time 
the Landau-Zener transition time for one of the crossings $t_{LZ}\approx
\Omega_{eff}(t_{i})/\alpha$ is of order 7, where $t_{i}$ is the crossing
time. All that is required is that the transition time, $t_{LZ}$, be less
than the time between crossings, $\Delta t$. It is clear that the direction
of population flow is controlled by the sign of the chirp. In our example of
positive chirp, population is transferred to the positive n-mode state 
(the $+25$th state, Fig.~2). By changing the chirp sign we are able to 
switch the direction of the population transfer and populate the $-25$th 
state at a later time. 

It is also possible to coherently split an atomic beam using a frequency 
chirped standing wave. One way to accomplish this task is to use an 
additional laser pulse of opposite chirp. The idea is to create 
simultaneously two frequency-chirped standing waves. A positively 
chirped standing wave provides a momentum transfer in the $+n \hbar k$ 
branch of momentum states while negatively chirped wave works in 
parallel on the $-n \hbar k$ branch.

In this case the modified equation for the ground state wave function in 
the momentum representation takes the form 
\begin{equation}
\begin{array}{lll}
i\dot{a}(p,t) = & \frac{p^{2}}{2m\hbar} a(p,t) - 
\Omega_e(t) a(p - 2 \hbar k, t) &  \\ 
& \, &  \\ 
& - \Omega_e^*(t) a(p + 2 \hbar k, t) \ , 
\end{array}
\label{eq:momsp3}
\end{equation}
where $\Omega_e(t) =\Omega_e^+(t) \exp[i \phi_{1}(t)]  + \Omega_e^- (t) 
\exp[i \phi_{2}(t)]$, $\phi_{1}(t) = \delta_{1}(t) - \delta_{2}(t) - 
\delta_0 t$, $\phi_{2}(t) = \delta_{1}(t) + \delta_{3}(t) + 
\delta_0^{\prime} t$, $\dot{\delta}_{3}(t)$ represents chirp of an 
additional pulse, $\Omega_{e}^{\pm}(t) = \Omega _{k}(t) 
\Omega_{-k}^{\pm}(t)/(4\Delta)$, $\Omega_{k}(t) = \mu_{12}E_{k}(t)/
\hbar $, $\Omega_{-k}^{\pm}(t) = \mu_{12}E_{-k}^{\pm}(t)/\hbar$. 
In Eq.~(\ref{eq:momsp3}) we have omitted the common light shift. 

Equation~(\ref{eq:momsp3}) can be rewritten as a linear system with the
Hamiltonian as in Eq.~(\ref{eq:Hmatrix}). However in this case the coupling
between momentum state wave functions is different for the negative and
positive branches: $\Omega_{-n}(t)=\Omega_{e}^{+}(t) + \Omega_{e}^{-}(t)\exp
\left\{ -i \delta_{23}^{\prime}(t) \right\}$, $\Omega_{+n}(t)=
\Omega_{e}^{+}(t)\exp \left\{ i \delta_{23}^{\prime}(t) \right\} + \Omega
_{e}^{-}(t)$, where $\delta_{23}^{\prime}(t) = \delta_{2}(t)+\delta_{3}(t) 
+\left[\delta_{0} + \delta_{0}^{\prime}\right] t$. 
More important is that the quasi-energies $E_{-n}(t)$ and $E_{+n}(t)$ are 
controlled by the different chirps, $\dot{\phi}_{1}(t)$ and $
\dot{\phi}_{2}(t)$: $E_{-n,+n}(t)= \hbar \left\{\omega_{k}n^{2} + 
n \dot{\phi}_{1,2}(t) \right\}$. As a result one can coherently control 
wave function dynamics in both branches of momentum states.
\begin{figure}[h]
\includegraphics[height=9cm,width=8cm]{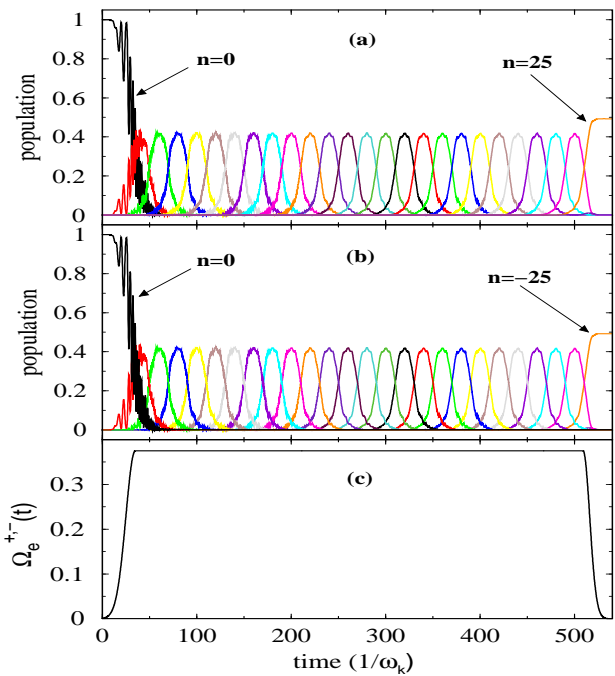}
\noindent 
\caption{Population dynamics for an atomic beam-splitter. 
We choose $t_c=20/\protect\omega_k$, 
$\protect\alpha=0.1 \protect\omega_k^2$. Solid lines 
show the population flow from initial state $n=0$ to the target state 
$n = 25$ - (a) and to the target state $n = -25$ - (b). 
(c) Shape of the laser pulse.}
\end{figure}

Figure~3 illustrates the dynamics for the case of an atom beam splitter. 
We choose $\dot{\phi}_{1}(t)=-\dot{\phi}_{2}(t)= \alpha (t-t_{c})$, 
$E_{k}(t)=E_{-k}^{+}(t)=E_{-k}^{-}(t)$, $\delta _{0}=
\delta _{0}^{\prime }=0$, and the chirp rate $\alpha =0.1\omega _{k}^{2}$. 
Our target states are $+50 \hbar k$ and $-50\hbar k$. At later time we 
have almost perfect splitting of the population between our target states. 
By adjusting the efficiency of the first avoided crossing taking place 
between the $0$ and $\pm 1$ states in such a way that the probability 
of a population transfer in both directions is equal we achieve 
symmetrical beam splitting. If the driving fields derived from the same 
laser, a coherent superposition of the $\pm 50 \hbar k$ states can be 
created. Note that the non-adiabatic couplings at the time of the first 
avoided crossings are responsible for small amount of population 
($\approx 0.5$\%) remaining in low lying levels. We note that
this scheme is very robust as well as selective and controllable. By
increasing or decreasing the pulse duration by 
$N\Delta t=N2\omega_{k}/\alpha$ ($N$ is an integer), one can create a 
beam splitter of larger or smaller angle.

Let us give preliminary values of the pulse parameters which can be used 
in experiments. All frequencies in our simulations have been normalized 
to the recoil frequency. Assuming that $\omega_k/2\pi$ is about 50 kHz 
we find that to make a beam splitter ($n=\pm 25$) one has to tune the 
frequency difference in the range of $50 \omega_k/2\pi = 2.5~MHz$. 
That means a transform limited pulse of $\approx$ 60~nsec duration should 
be used to produce the chirp $\alpha/2\pi = 0.1~\omega_k^2 \approx 
1.6~kHz/\mu sec$ (see Fig.~2~and~3).

One method of producing appropriate chirp rates is to use acousto-optical
modulators (AOM) as in studies of Landau-Zener tunneling \cite{raizen}. 
In this case one must devise a method to transfer the temporal frequency 
chirp to a spatial one as an atomic beam passes through the interaction 
zone. One possibility for producing a spatial chirp directly is to use 
the Doppler shift associated with curved wave 
fronts~\cite{KroPRA85,EksOCM96}.

If a Bragg beam splitter of this type is to be used as an element of an 
atom interferometer, the transverse momentum spread of the atomic beam 
must be less than some critical $\Delta p_{c}$ of order $\hbar k/\left( 
2n\omega_{k}T\right) $ where $T$ is the total pulse duration. 
Preselection of this narrow velocity range by a preparation pulse may 
be needed. This severe restriction on the momentum spread is based on 
the assumption that the different momentum components in the beam are 
uncorrelated. However, if a coherent source such as a BEC is used, 
our wave packet simulations show that our schemes are valid even for a 
transverse momentum spread as large as $0.5\hbar k$.

We express our appreciation for many useful discussions with 
P.~Bucksbaum, G.~Raithel, B.~Dubetsky. This work is supported by the 
U. S. Office of Army Research under Grant No. DAAD19-00-1-0412 the 
Michigan Center for Theoretical Physics, and the National Science 
Foundation under Grant No. PHY-9800981, Grant No. PHY-0098016, and 
the FOCUS Center Grant.


\begin{references}
\bibitem{GusPRL97}  
T.~L.~Gustavson, P.~Bouyer, and M.~A.~Kasevich,
Phys.~Rev.~Lett., {\bf 78}, 2046 (1997).

\bibitem{ZobPRA99}  
O.~Zobay, E.~V.~Goldstein, and P.~Meystre, Phys.~Rev.~A, 
{\bf 60}, 3999 (1999).

\bibitem{FujPRL00}  
J.~Fujita, S.~Mitake, and F.~Shimizu, Phys.~Rev.~Lett., 
{\bf 84}, 4027 (2000).

\bibitem{BrePRL99}  
G.~K.~Brennen {\it et al.}, Phys.~Rev.~Lett., {\bf 82}, 1060 (1999).

\bibitem{JakPRL99}  
D.~Jaksch {\it et al.}, Phys.~Rev.~Lett., {\bf 82}, 1975 (1999).

\bibitem{BerAI}  
See, for example, {\it Atom Interferometry}, edited by
P.~R.~Berman, (Academic, Cambridge, MA, 1997).

\bibitem{AdaPRS94}  
C.~S.~Adams, M.~Sigel, and J.~Mlynek, Phys.~Reports {\bf 240}, 143 (1994).

\bibitem{MetPRS94}  
H.~Metcalf and P.~van~der~Straten, Phys.~Reports {\bf 244}, 203 (1994).

\bibitem{WalPRS95}  
H.~Wallis, Phys.~Reports {\bf 255}, 203 (1995).

\bibitem{KasPRL91}  
M.~Kasevich and S.~Chu, Phys.~Rev.~Lett. {\bf 67}, 181 (1991).

\bibitem{McgPRL00}  
J.~M.~McGuirk, M.~J.~Snadden, and M.~A.~Kasevich,
Phys.~Rev.~Lett. {\bf 85}, 4498 (2000).

\bibitem{DubPRA01}  
For a review of large angle beam splitters see B.~Dubetsky 
and P.~R.~Berman, Phys.~Rev.~A, {\bf 64}, 063612 (2001).

\bibitem{BerPRA97}  
P.~R.~Berman, B.~Bian, Phys.~Rev.~A, {\bf 55}, 4382 (1997).

\bibitem{MarPRL91}  
P.~Marte {\it et al.}, Phys.~Rev.~A, {\bf 44}, R4118 (1991).

\bibitem{BanPRA93}  
Y.~B.~Band, Phys.~Rev.~A, {\bf 47}, 4970 (1993).

\bibitem{GauCPL88}  
U.~Gaubatz {\it et al.}, Chem.~Phys.~Lett., {\bf 149}, 463 (1988).

\bibitem{PilPRA93}  
P.~Pillet {\it et al.}, Phys.~Rev.~A, {\bf 48}, 845 (1993).

\bibitem{WeiPRL94}  
M.~Weitz, B.~C.~Young, and S.~Chu, Phys.~Rev.~Lett., {\bf 73}, 2563 (1994).

\bibitem{LawPRL94}  
J.~Lawall, M.~Prentiss, Phys.~Rev.~Lett., {\bf 72}, 993(1994).

\bibitem{GolPRL94}  
L.~S.~Goldner {\it et al.}, Phys.~Rev.~Lett., {\bf 72}, 997 (1994).

\bibitem{raizen}  
S.~R.~Wilkinson {\it et al.}, Phys.~Rev.~Lett., {\bf 76},
4512 (1996); S.~R.~Wilkinson {\it et al.}, Nature, {\bf 387}, 575 (1997).

\bibitem{Che95R3417}  
S.~Chelkowski and G.~Gibson, Phys.~Rev.~A {\bf 52},
R3417 (1995); J.~C.~Davis and W.~S.~Warren, J.~Chem.~Phys., {\bf 110}, 
4229 (1999).

\bibitem{ChaCPL373}  
B.~Y.~Chang {\it et al.}, Chem.~Phys.~Lett., {\bf 341}, 373 (2001).

\bibitem{PeiPRA2989}  
E.~Peik {\it et al.}, Phys.~Rev.~A, {\bf 55}, 2989 (1997).

\bibitem{KroPRA85}  
J.~P.~C.~Kroon {\it et al.}, Phys.~Rev.~A, {\bf 31}, 3724 (1985).

\bibitem{EksOCM96}  
C.~R.~Ekstrom {\it et al.}, Opt.~Commun., {\bf 123}, 505 (1996).
\end{references}
\end{document}